\documentclass[11pt,english]{elsarticle}
\usepackage{mathptmx}

\usepackage[T1]{fontenc}
\usepackage[latin9]{inputenc}
\usepackage[letterpaper]{geometry}
\usepackage{color}
\usepackage{amsmath}
\usepackage{amssymb}
\usepackage{graphicx}
\usepackage{esint}

\makeatletter

\newcommand{\lyxdot}{.}

\@ifundefined{showcaptionsetup}{}{%
 \PassOptionsToPackage{caption=false}{subfig}}
\usepackage{subfig}
\makeatother

\usepackage{babel}
\begin{document}

\title{Finite spatial-grid effects in energy-conserving particle-in-cell
algorithms}

\author[coronado]{D. C. Barnes\corref{cor1}}

\ead{coronadocon@msn.com}

\author[lanl]{L. Chacón}

\cortext[cor1]{Corresponding author}

\address[coronado]{Coronado Consulting, Lamy, NM 87540}

\address[lanl]{Los Alamos National Laboratory, Los Alamos, NM 87545}

\address{}
\begin{abstract}
Finite-grid (or aliasing) instabilities are pervasive in particle-in-cell
(PIC) plasma simulation algorithms, and force the modeler to resolve
the smallest (Debye) length scale in the problem regardless of dynamical
relevance. These instabilities originate in the aliasing of interpolation
errors between mesh quantities and particles (which live in the space-time
continuum). Recently, strictly energy-conserving PIC (EC-PIC) algorithms
have been developed that promise enhanced robustness against aliasing
instabilities. In this study, we confirm by analysis that EC-PIC is
stable against aliasing instabilities for stationary plasmas. For
drifting plasmas, we demonstrate by analysis and numerical experiments
that, while EC-PIC algorithms are not free from these instabilities
in principle, they feature a benign stability threshold for finite-temperature
plasmas that make them usable in practice for a large class of problems
(featuring ambipolarity and realistic ion-electron mass ratios) without
the need to resolve Debye lengths spatially. We also demonstrate that
this threshold is absent for the popular momentum-conserving PIC algorithms,
which are therefore unstable for both drifting and stationary plasmas.
\end{abstract}
\maketitle

\section{Introduction}

The Particle-In-Cell (PIC) method has proven to be one of the most
useful tools for simulation of plasmas. In addition to its intuitive
feel of reproducing collective behavior by following individual numerical
markers, it lends itself easily to situations in which complex boundary
conditions, and/or interaction between several species (such as neutrals
with ionization and related physics) must be accommodated. Along with
these advantages, traditional PIC methods manifest some quite stringent
limitations. Modern algorithmic improvements have worked toward eliminating
or controlling these limitations. One such limitation is associated
with the aliasing or finite-grid instability, first studied systematically
by Langdon \citep{langdon1970spatialeffects} and later by Birdsall
and Maron \citep{birdsall1980plasma}. The classic book by Birdsall
and Langdon \citep{birdsall-langdon} contains a useful introduction
to this subject in Chapter 8, and our present work follows closely
this approach. 

There has been renewed interest on the subject of aliasing instabilities
in PIC methods. A recent discussion from a theoretical and empirical
viewpoint \citep{brackbill2016energy} shows that this instability
is ubiquitous in PIC methods. The study in Ref. \citep{meyers2015numerical}
derives dispersion relations (verified by numerical experiments) for
the aliasing instabilities in a classical electromagnetic explicit
algorithm, with momentum-conserving deposition. The importance of
the lack of spectral fidelity of the particle representation in aliasing
instabilities in explicit, momentum-conserving PIC is explored in
Ref. \citep{huang2016finite}.

The motivation for this study is the apparent lack of aliasing instabilities
in recently proposed fully implicit, energy-conserving PIC algorithms
\citep{chen-jcp-11-ipic,markidis2011energy,taitano-sisc-13-ipic,chen-cpc-14-darwin,chen-cpc-15-darwin2d},
which take advantage of modern iterative solvers, e.g. Jacobian-free
Newton-Krylov (JFNK) methods \citep{knoll-jcp-04-nk_rev} or Picard-accelerated
methods \citep{walker2011anderson}. This empirical observation has
been reported by all groups investigating this approach. Hence the
situation is unclear, as on the one hand, theory clearly indicates
the possibility of instability for cold, drifting beams \citep{langdon-jcp-73-ec_vs_mc},
and on the other hand, it is difficult to observe any difficulties
associated with these instabilities in practical applications.

In usual PIC algorithms, the Debye length must equal or exceed the
cell size for stability against the aliasing instability, even though
the physics often does not depend on Debye-length physics. For example,
dense plasmas often are described well by the quasi-neutral approximation
in most of the domain, so that fidelity to the relevant physics could
be obtained with much larger linear cell size. The impact of using
a cell size compatible with the physics requirement vs. one of the
Debye length is enormous in higher dimensions, and can easily lead
to a one million times reduction of computational expense in both
time and memory. Thus, there is a strong motivation for better understanding
of this subject.

The approach of this study is to understand by numerical analysis
and corresponding numerical experiments the effect of finite spatial
grid effects for both cold and thermal drifting populations on the
stability of energy-conserving PIC (EC-PIC) algorithms. Given that
finite-grid instabilities are electrostatic in nature, we focus the
discussion on electrostatic EC-PIC algorithms (such as the one in
Ref. \citep{chen-jcp-11-ipic}, which is also implicit and charge
conserving, and we term here CCB). Given the focus on finite-grid
instabilities, we consider the $\Delta t\rightarrow0$ limit of CCB.
The assumption of zero time step means that any conclusions can be
applied to an energy-conserving PIC method with any time-stepping
scheme, and this feature is discussed in our conclusions.

Energy-conserving PIC methods were introduced by Lewis and others
around 1970 \citep{lewis1970energy} and shortly afterward were examined
theoretically and empirically for aliasing instabilities \citep{langdon-jcp-73-ec_vs_mc}.
This early work indicated that there was no particular distinction
of EC-PIC approaches compared with the established momentum-conserving
methods, as far as avoidance of aliasing instabilities. It was shown
that: 1) energy is not particularly well-conserved by the early, time-explicit
implementations; 2) cold-drifting beams were unstable with comparable
linear growth rates and comparable nonlinear consequences; 3) lack
of momentum conservation allowed conversion of drift energy to thermal
energy \emph{via} the aliasing instability. While some hopeful signs
that aliasing instabilities might be avoided in some applications,\footnote{See Ref. \citep{birdsall-langdon}, p. 177 comment in caption of Fig.
8-12b concerning the stability threshold on the drift velocity for
energy-conserving methods.} the situation seemed not sufficiently hopeful to advance EC-PIC approaches
instead of momentum-conserving ones.

In the last nearly half century, computing has advanced beyond all
expectations, and this explosion in capability now allows detailed
examination of a sufficiently large parameter space. Quite different
conclusions can be drawn from this enlarged perspective. We find a
new stability regime for drifting thermal (Maxwellian) electrons,
which is relevant for modern applications with realistic mass ratios.
In fact, this regime may be (and has been) exploited in applying PIC
to dense, quasi-neutral simulations, completely avoiding aliasing
instabilities with cell sizes up to more than 100 times plasma Debye
length. We follow the established approach to derive and solve (usual)
dispersion relations (DRs) associated with spatial grid effects and
to compare growth-rate predictions with numerical experiments. Our
extended survey of these results leads to new and hopeful conclusions
about aliasing in energy-conserving methods and explains the apparent
success of these methods in avoiding Debye length restrictions. In
fact, while EC-PIC does not suffer from aliasing instabilities in
stationary thermal populations (as reported for instance in Ref. \citep{lapenta-jcp-17-ec_ipic}),
aliasing instabilities are present in EC-PIC algorithms for cold-beam
drifting populations. While the combination of smoother (continuously
differentiable) charge-conserving scatter and binomial smoothing (as
proposed in the original CCB algorithm \citep{chen-jcp-11-ipic})
is useful in significantly reducing growth rates of linearly unstable
cold-beam aliasing modes, and in ameliorating the Debye-length resolution
requirement for stability, these are not required to access a previously
unidentified stability region, namely, EC-PIC algorithms are stable
below a critical Mach number (defined as the ratio of the electron
drift to thermal velocities) of $\mathcal{O}(1)$. The implication
is that, as we discuss, EC-PIC algorithms may not require resolution
of the Debye length in ambipolar plasma systems when using realistic
ion-electron mass ratios, and aliasing instability may be easily avoided
in practical applications.

The rest of this paper is organized as follows. Section \ref{sec:dispersion-relations}
derives the relevant dispersion relations for the numerical scheme
for cold (zero temperature) and warm beams. The numerical analysis
of the dispersion relations to find unstable roots in performed in
Sec. \ref{sec:numerical_analysis}. Numerical experiments verifying
key results from the analysis and illustrating the main conclusions
of this paper are provided in Sec. \ref{sec:numerical_exp}. Finally,
we conclude in Sec. \ref{sec:conclusions}.

\section{Dispersion relations}

\label{sec:dispersion-relations}

In this section, we derive the required dispersion relations for cold
and warm beam cases, including binomial smoothing and all effects
of a discrete spatial grid (but for the limit of zero time step).

\subsection{Equivalent electrostatic formulation}

\label{subsec:EqES}

An exact charge-conserving scheme such as CCB is mathematically equivalent
to a formulation in terms of the electrostatic potential $\phi$.
Indeed charge conservation means that there exists a density shape
function $s$ which is related to the current/electric field shape
function $\hat{s}$ by finite differences to round-off accuracy. Thus,
the edge-centered $\hat{s}$ and the vertex-centered $s$ satisfy
\citep{chacon-jcp-13-ipic_curv}

\begin{equation}
s_{i}\left(x^{1}\right)-s_{i}\left(x^{0}\right)=-\left(x^{1}-x^{0}\right)\frac{\hat{s}_{i+1/2}\left(x^{1/2}\right)-\hat{s}_{i-1/2}\left(x^{1/2}\right)}{\Delta}\label{eq:ccshapes}
\end{equation}

\noindent where $\Delta$ is the mesh spacing (assumed uniform here).

Multiplying this by the vertex-centered potential $\phi_{i}$ and
summing over $i$, we obtain, after re-arranging the sum on the right

\begin{equation}
-\frac{\phi\left(x^{1}\right)-\phi\left(x^{0}\right)}{x^{1}-x^{0}}=\sum_{i}\hat{s}_{i+1/2}\left(x^{1/2}\right)\left(-\frac{\phi_{i+1}-\phi_{i}}{\Delta}\right)\label{eq:DQ}
\end{equation}

\noindent where we have defined the interpolation of the potential
as $\phi\left(x\right)=\sum_{i}s_{i}\left(x\right)\phi_{i}$. The
right hand side of Eq. (\ref{eq:DQ}) is just the energy-conserving
interpolation of the edge-centered electric field used in CCB, where
the electric field is defined as $E_{i+1/2}=-\left(\phi_{i+1}-\phi_{i}\right)/\Delta$.

\subsection{Cold drifting beam}

\label{subsec:cold-beam}

The results of the previous sub-section shows that the CCB orbit equations
may also be written as
\begin{flalign}
x^{n+1}-x^{n} & =\Delta tv^{n+1/2}\nonumber \\
v^{n+1}-v^{n} & =-\frac{q_{p}\Delta t}{m_{p}}\frac{\phi^{n+1/2}\left(x^{n+1}\right)-\phi^{n+1/2}\left(x^{n}\right)}{x^{n+1}-x^{n}}\label{eq:FD-CCB}
\end{flalign}
where $\phi\left(x\right)=\sum_{i}s\left(x-x_{i}\right)\phi_{i}$
is the potential interpolated from the mesh, and $s(x-x_{i})$ is
the density shape function. It is obvious that, in the limit of zero
time step, this yields:
\begin{flalign*}
\dot{x} & =v\\
\dot{v} & =-\frac{q_{p}}{m_{p}}\partial_{x}\phi\left(x\right)
\end{flalign*}
The consistent charge deposition associated with this scheme is

\begin{equation}
\rho_{i}=\frac{1}{\Delta}\sum q_{p}s\left(x-x_{i}\right),\label{eq:Ndep}
\end{equation}
with $\Delta$ the grid spacing. In the small-amplitude limit, the
particle positions $x$ are displaced by $\delta x$ and the perturbed
charge density reads:
\begin{equation}
\rho_{i}=\frac{1}{\Delta}\sum_{p}q_{p}\delta x\partial_{x}s\left(x-x_{i}\right).\label{eq:Linear-ro}
\end{equation}
In this way, the only spline which enters the formulation is $s$,
with (infinite) Fourier transform $S\left(k\right)$. Choosing $s$
to be the $m$-th order B-spline, we have:
\begin{equation}
S(k)=\left[\frac{\sin\left(k\Delta/2\right)}{k\Delta/2}\right]^{m+1}.\label{eq:shapef-fourier}
\end{equation}

Following the aliasing formulation given in Birdsall and Langdon \citep{birdsall-langdon},
we find the DR for a beam with drift velocity $U$ as:

\begin{equation}
D=1-\frac{\omega_{p}^{^{2}}}{K^{2}}\sum_{q}\frac{k_{q}^{2}\left[S\left(k_{q}\right)\right]^{2}}{\left(\omega-k_{q}v\right)^{2}}=0,\label{eq:cold-dispersion-relation}
\end{equation}
where $k_{q}=k+2\pi q/\Delta$ is the alias wave number, and

\begin{equation}
K^{2}=\left[k\frac{\sin\left(k\Delta/2\right)}{k\Delta/2}\right]^{2}\label{eq:K2}
\end{equation}
is the negative of transform of the finite-difference Laplacian. Equation
\ref{eq:cold-dispersion-relation} is identical to the one found for
Lewis' energy-conserving method \citep{lewis1970energy} in Sec. 8-11
of Ref. \citep{birdsall-langdon}. Using Eqs. \ref{eq:shapef-fourier}
and \ref{eq:K2}, we find: 
\begin{equation}
D=1-\omega_{p}^{2}\left[\sin\left(k\Delta/2\right)\right]^{2m}\cos^{4}\left(k\Delta/2\right)\sum_{q}\frac{1}{\left(k_{q}\Delta/2\right)^{2m}\left(\omega-k_{q}U\right)^{2}}=0,\label{eq:dispersion_relation}
\end{equation}
where the $\cos^{4}\left(k\Delta/2\right)$ factor is present when
a binomial $\left(1,2,1\right)/4$ filter is applied on the mesh twice
per cycle, as originally proposed in CCB \citep{chen-jcp-11-ipic}. 

We introduce the normalized variables $\kappa=k\Delta$, $u=U/\omega_{p}\Delta$
and $\bar{\omega}=\omega/\omega_{p}$ to re-write Eq. \ref{eq:dispersion_relation}
as:
\begin{equation}
D=1-\left[\sin\left(\kappa/2\right)\right]^{2m}\cos^{4}\left(\kappa/2\right)\sum_{q}\frac{1}{\left(\kappa/2+q\pi\right)^{2m}\left(\bar{\omega}-\kappa u-2q\pi u\right)^{2}}=0.\label{eq:DR-cold}
\end{equation}

The case $m=0$ with no digital filter, while unphysical (one cannot
interpolate $\phi$ with nearest-grid-point formulae, as they cannot
capture electric fields correctly), is particularly simple, because
of the identity
\[
\left(\pi\csc\pi z\right)^{2}=\sum_{q}\frac{1}{\left(z-q\right)^{2}}.
\]
Thus, for this case, the dispersion relation becomes:
\begin{equation}
1=\frac{\left(2u\right)^{-2}}{\sin^{2}\left[\left(2u\right)^{-1}\left(\bar{\omega}-\kappa u\right)\right]}\label{eq:l1dispersion}
\end{equation}
It is easy to see from Eq. \ref{eq:l1dispersion} that there is stability
for $u>1/2$ and instability otherwise (simply using the fact that
$\left|\sin(x)\right|\leq1$ for real $x$). This suggests that energy
conserving PIC may be unstable to beam modes when the drift velocity
is small enough, just as in explicit PIC methods. We will consider
the realistic case of $m>0$ numerically later in this study.

\subsection{Warm beam}

\label{subsec:warm-beam}

The previous results can be extended to the case of a drifting Maxwellian,

\begin{equation}
\bar{f}=\frac{\bar{n}e^{^{-m\left(v-U\right)^{2}/2T}}}{\sqrt{2\pi T/m}}.\label{eq:Max}
\end{equation}
The solution for the dispersion relation can be obtained by convolving
the cold-beam solution Eq. \ref{eq:cold-dispersion-relation} with
the equilibrium distribution function $\bar{f}$, to find:

\begin{equation}
1-\frac{\omega_{p}^{^{2}}}{K^{2}}\sum_{q}k_{q}^{2}\left[S\left(k_{q}\right)\right]^{2}\int_{-\infty}^{\infty}dv\frac{e^{-v^{2}/2v_{T}^{2}}}{\sqrt{2\pi}v_{T}\left(\omega-k_{q}v\right)^{2}}=0,\label{eq:warm-dispersion-relation}
\end{equation}
where $v_{T}^{2}=T/m$. There results:
\begin{equation}
1+4\frac{\left[\sin\left(\kappa/2\right)\right]^{2m}}{\lambda^{2}}\sum_{q}\frac{1+\Omega_{q}Z\left(\Omega_{q}\right)}{\left(\kappa/2+\pi q\right)^{2\left(m+1\right)}}=0,\label{eq:DR2}
\end{equation}
where the dimensionless argument of the Fried-Conte function $Z$
is:

\begin{equation}
\Omega_{q}=\frac{1}{\sqrt{2}\lambda}\left(\frac{\bar{\omega}}{\kappa+2\pi q}-u\right)\mathrm{sign}\left(\kappa+2\pi q\right),\label{eq:Omq}
\end{equation}
with $\lambda=\lambda_{D}/\Delta$. The sign function in Eq. \ref{eq:Omq}
is required so that the imaginary part of $\omega$ is treated properly
in the $Z$ functions.

As in the cold-beam case, the effects of the $\left(1,2,1\right)/4$
binomial mesh filter can be readily included with an additional $\cos^{4}\left(k\Delta/2\right)$
factor, and the final dispersion relation reads:

\begin{equation}
1+4\frac{\left[\sin\left(\kappa/2\right)\right]^{2m}\left[\cos\left(\kappa/2\right)\right]^{4}}{\lambda^{2}}\sum_{q}\frac{1+\Omega_{q}Z\left(\Omega_{q}\right)}{\left(\kappa/2+\pi q\right)^{2\left(m+1\right)}}=0.\label{eq:DR2-binomial}
\end{equation}

We show next that Eq. \ref{eq:DR2-binomial} reduces to the cold-beam
result when $\lambda$ goes to zero. In this case, the arguments of
the $Z$ functions become large and we use the asymptotic limits

\begin{equation}
Z\left(\varsigma\right)\sim-\frac{1}{\varsigma}\left(1+\frac{1}{2\varsigma^{2}}+\ldots\right)+\mathcal{O}\left(e^{-\varsigma^{2}/2}\right)\label{eq:Zasm}
\end{equation}
\textcolor{black}{The last (exponential) term will not be important}\textcolor{red}{{}
}because we focus here only on growing modes for which this term is
exponentially small. Substituting this into Eq. \ref{eq:DR2-binomial}
we find, to leading order:

\begin{equation}
1-4\frac{\left[\sin\left(\kappa/2\right)\right]^{2m}\left[\cos\left(\kappa/2\right)\right]^{4}}{\lambda^{2}}\sum_{q}\frac{1/2\Omega_{q}^{2}}{\left(\kappa/2+\pi q\right)^{2\left(m+1\right)}}=0,\label{eq:DR-3}
\end{equation}
which, when substituting Eq. \ref{eq:Omq}, gives the cold-beam dispersion
relation of Eq. \ref{eq:DR-cold}.

\section{Numerical analysis}

\label{sec:numerical_analysis}

\begin{figure}
\begin{centering}
\includegraphics[width=0.48\columnwidth]{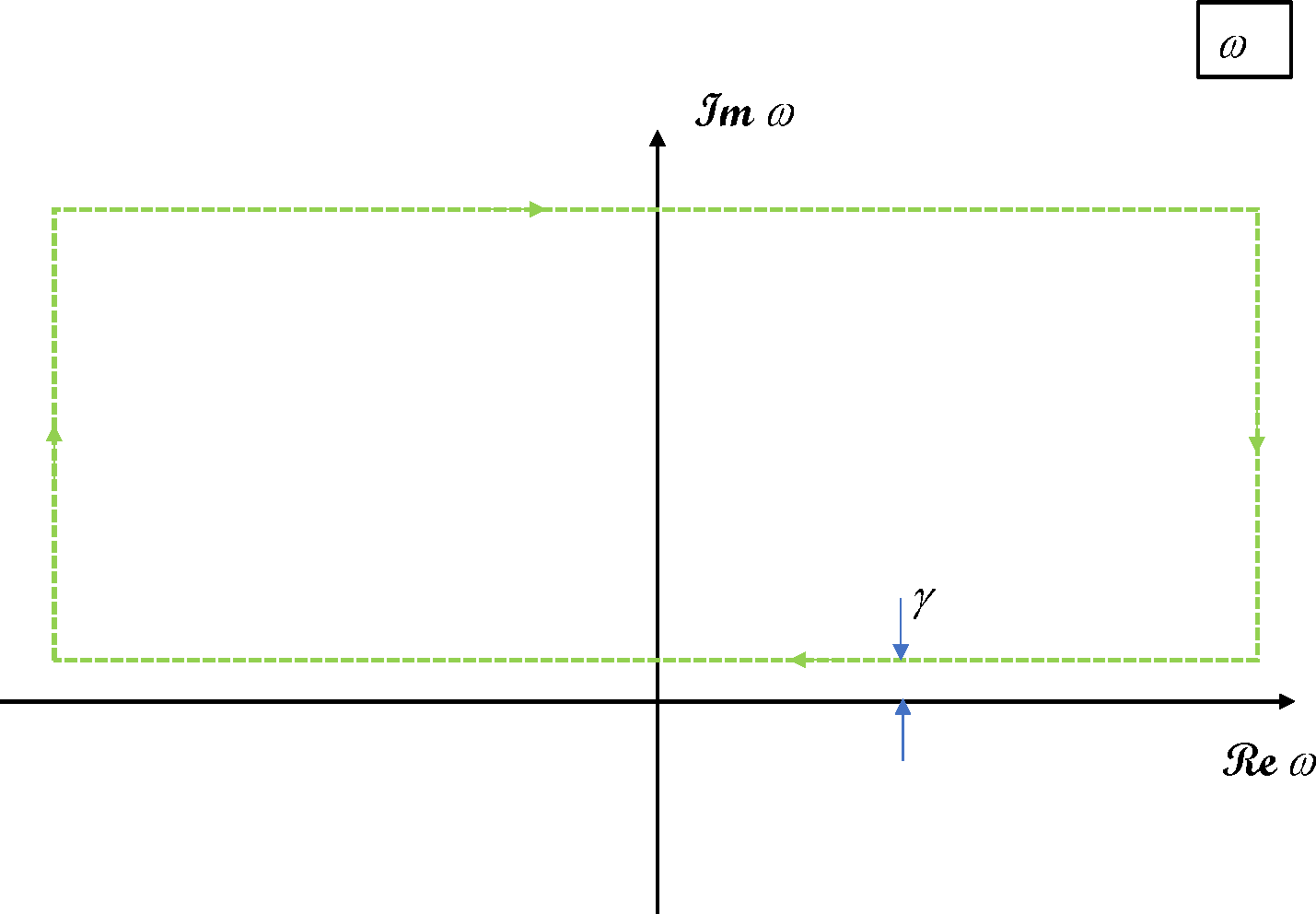}
\par\end{centering}
\caption{\label{fig:box}Integration box for finding unstable complex roots.}
\end{figure}

The dispersion relation is studied by specifying the dimensionless
parameters $\lambda,u$ along with $m$ and an artificial limit $Q$
on the sum on $q$. The normalized wave number $\kappa$ is scanned
over the first Brillouin zone from $\left[-\pi,\pi\right]$. In order
to assure, as completely as possible, that all relevant roots are
located, we proceed in a systematic manner. A ``box'' in the complex
plane is defined as shown in Fig. \ref{fig:box}. The upper limit
of the box is chosen larger than any root (roots cannot have a growth
rate more than several $\omega_{p}$, so normalized $\bar{\omega}$
is limited to a few). Similarly, the left and right limits are chosen
so that all real frequencies of interest fall within the box. The
lower limit $\gamma$ is scanned from some small value (defined as
that for which the numerical evaluation of required functions is well
behaved) in small steps upward. At each step, the contour integral

\begin{equation}
I\left(\gamma\right)=\oint\frac{dz}{\epsilon}\label{eq:box}
\end{equation}

\noindent is evaluated, where $\epsilon=0$ is the dispersion relation
of interest. The lower limit is then raised in small steps, $I$ is
evaluated at each step using the LSODE ODE package \citep{lsode}
with adaptive, higher-order method, and its absolute value is examined.
When $\gamma$ exceeds the growth rate of the fastest-growing mode,
there are no singularities of the integrand located in the resulting
box, and the value of $I$ drops to near zero. The fastest-growing
mode is then determined to have a growth rate between the value of
$\gamma$ which gives a significant value of $I$ and that which gives
a near zero value.

The results of this procedure have been checked by an alternative
solution of the cold-beam dispersion relation of Sec. \ref{subsec:cold-beam}.
This may be formed into the problem of finding the zeros of a polynomial
of moderate degree (typically 28 to 40). Since such a problem has
a known number of roots, we use previously described methods \citep{bini1996numerical}
to find all zeros, along with error estimates of the location of each
root. These approximations are then used as starting points for the
secant method to find zeros of the DR as written in Eq. \ref{eq:DR-cold}.
Convergence of this procedure assures that no spurious roots are introduced
by conversion into a polynomial equation.

To recast the DR of Eq. \ref{eq:DR-cold} into the standard problem
of finding the zeros of a polynomial, we clear denominators
\begin{equation}
\prod_{q}\left(\bar{\omega}-\kappa u-2q\pi u\right)^{2}-\left[\sin\left(\kappa/2\right)\right]^{2m}\cos^{4}\left(\kappa/2\right)\sum_{q}\frac{1}{\left(\kappa/2+q\pi\right)^{2m}}\prod_{q'\neq q}\left(\bar{\omega}-\kappa u-2q'\pi u\right)^{2}=0\label{eq:DR-cold-poly}
\end{equation}
resulting in a polynomial equation of degree $2\left(2Q-1\right)$.
The Culham Laboratory-originated Fortran routine \texttt{PA16AD},
based on the work of Bini \citep{bini1996numerical}, is used to locate
the roots of this polynomial equation. Results agree with the box
method described previously, and show instability for a large range
of drift velocity and for every method considered here.

\begin{figure}
\begin{centering}
\includegraphics[width=0.48\columnwidth]{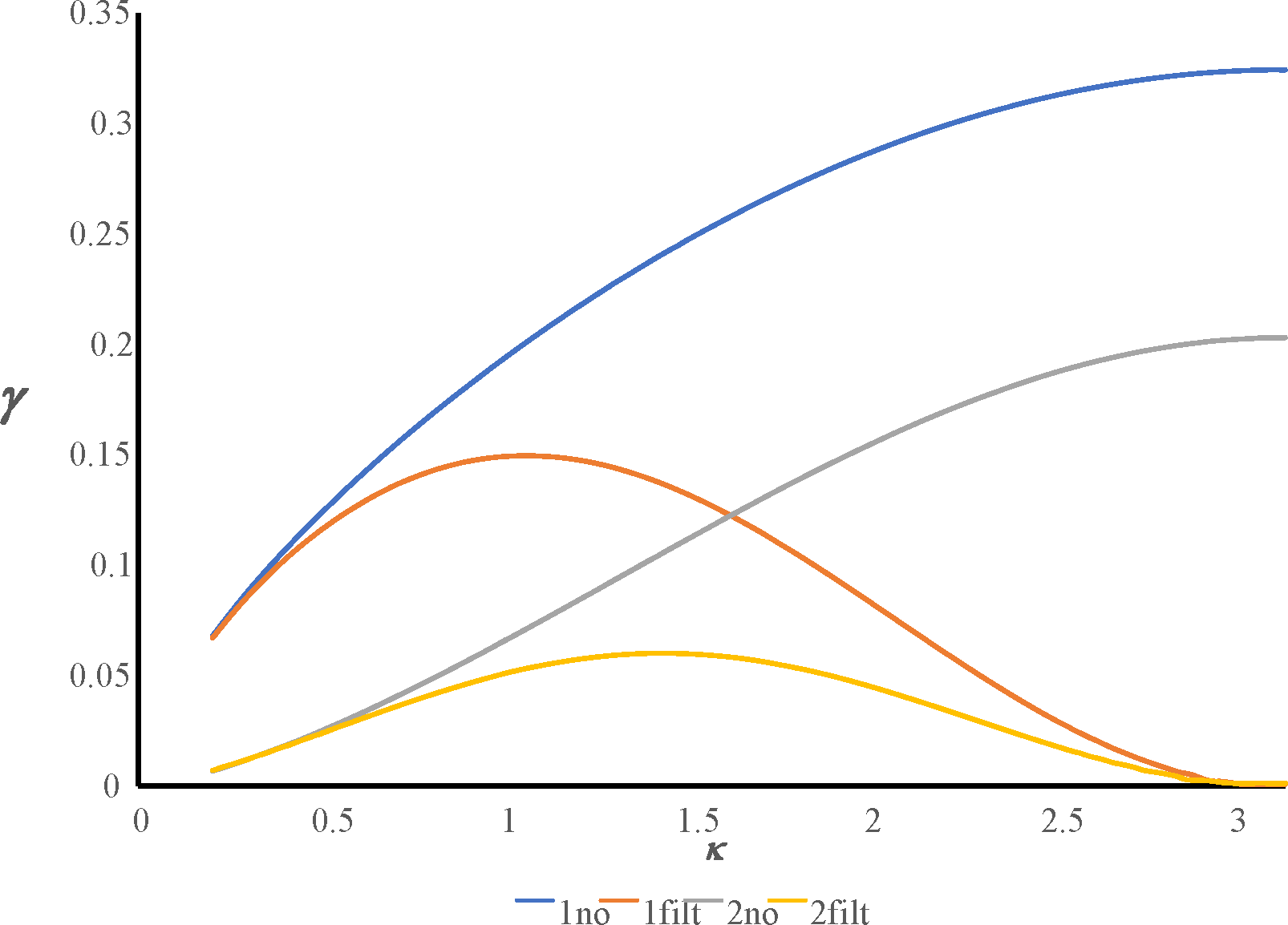}
\par\end{centering}
\caption{\label{fig:gam-v-k-4method}Cold-beam growth rate variation with wavenumber
for four choices of shape function ($m=1,2)$ and filtering (with
and without filtering), as follows: yellow ($m=2,$ with filtering),
gray ($m=2$, without filtering), red ($m=1,$ with filtering), and
blue ($m=1,$ without filtering).}
\end{figure}

A comparison of the growth rate for different choices of B-spline
order $m$ and filtering is shown in Fig. \ref{fig:gam-v-k-4method},
where the maximum (over $u$) growth rate is plotted \emph{vs. }$\kappa$
for the cases $m=1$ (linear spline) no binomial filter, $m=1$ with
binomial filter, $m=2$ (quadratic spline) no binomial filter, and
$m=2$ with binomial filter. As can be seen, it is possible to reduce
the linear growth rate by more than 5 times by use of smoother particle
shape and binomial filtering, with the combination of both providing
the most reduction.

\begin{figure}
\begin{centering}
\includegraphics[viewport=0bp 0bp 792bp 612bp,width=0.4\columnwidth]{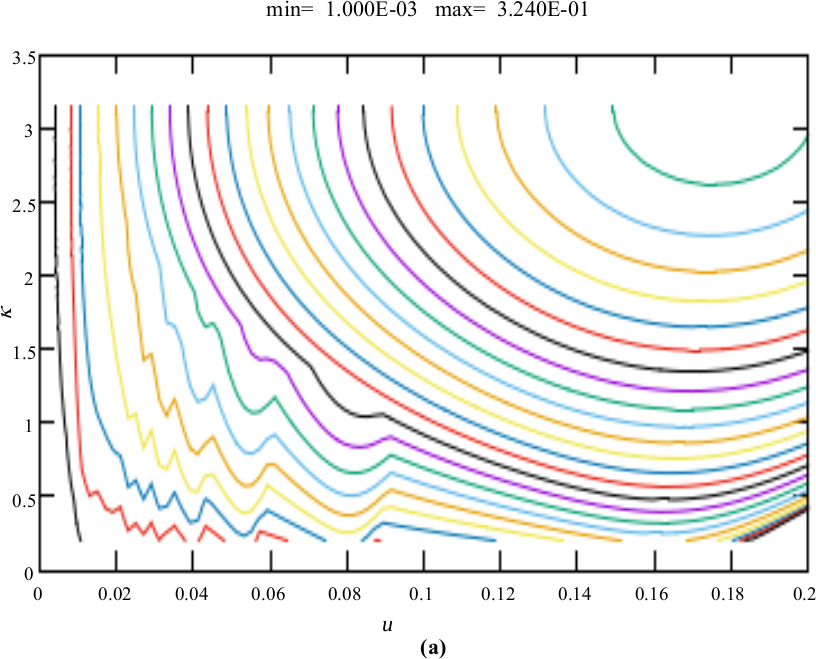}\includegraphics[viewport=0bp 0bp 792bp 612bp,width=0.4\columnwidth]{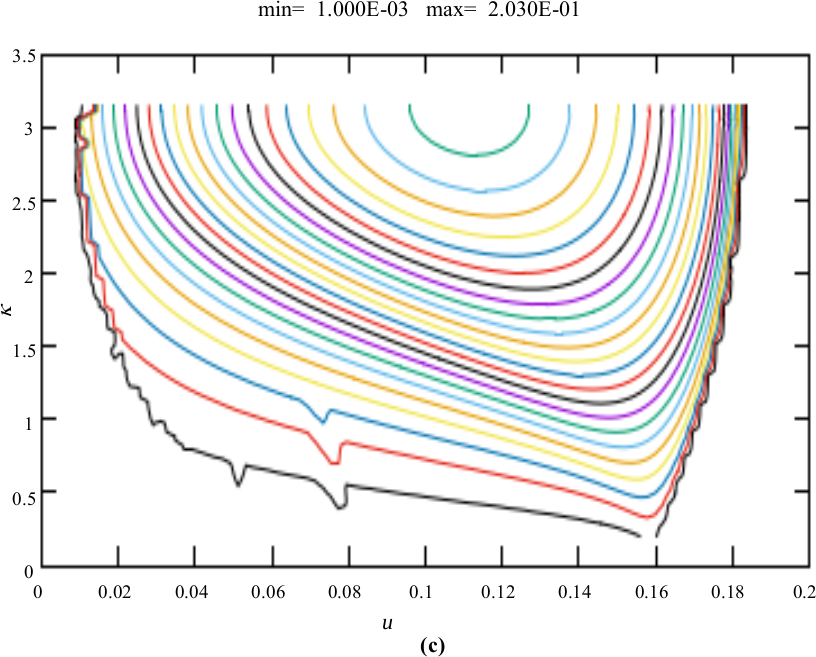}
\par\end{centering}
\begin{centering}
\includegraphics[viewport=0bp 0bp 792bp 612bp,width=0.4\columnwidth]{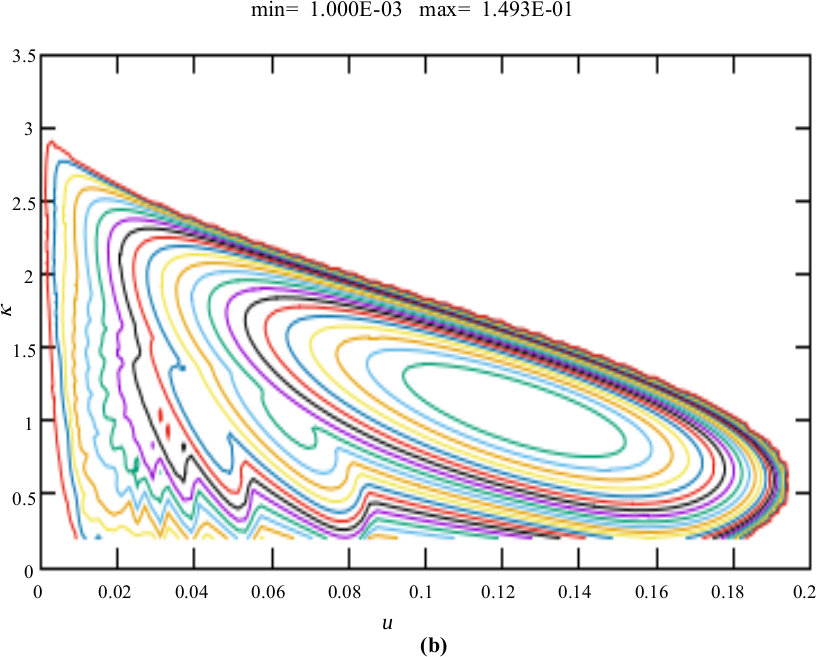}\includegraphics[viewport=0bp 0bp 792bp 612bp,width=0.4\columnwidth]{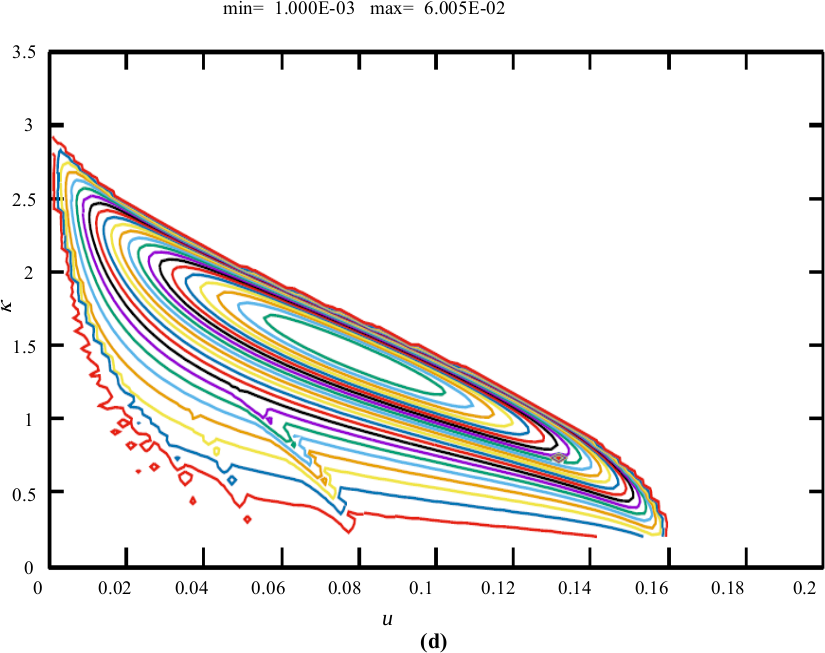}
\par\end{centering}
\caption{\label{fig:u-k contours-4methods}Contours of cold-beam growth rate
\emph{vs.} $u$ and $\kappa$ for four cases: (a) $m=1$ (linear spline)
no filter; (b) $m=2$ (quadratic spline) no filter; (c) $m=1$ with
filter; (d) $m=2$ with filter.}
\end{figure}

\begin{figure}
\begin{centering}
\includegraphics[width=0.48\columnwidth]{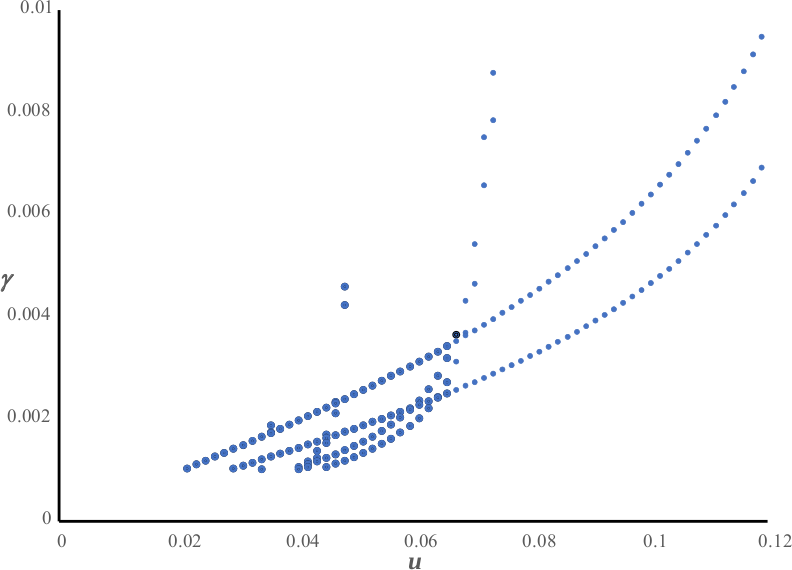}
\par\end{centering}
\caption{\label{fig:gamvuslice}Growth rate \emph{vs.} $u$ for all unstable
modes for $\kappa=0.497$ showing the existence of ridges at which
unstable modes become suddenly stable.}
\end{figure}

A more comprehensive picture of the cold-beam alias modes is provided
by Fig. \ref{fig:u-k contours-4methods}, where contours of growth
rate vs. both $u$ and $\kappa$ are displayed for each method.
 As can be seen the growth rates are significant (up to 32\% of $\omega_{p}$)
and ubiquitous, as pointed out in numerous places in the literature.
The ``ridges'' appearing in these contours result from multiple
branches of the dispersion relation solutions. Figure \ref{fig:gamvuslice}
shows all roots (obtained by the polynomial root finder method) \emph{vs.}
$u$ for a slice along a fixed $\kappa=0.497$ of the case of Fig.
\ref{fig:u-k contours-4methods} (d). The two prominent ridges near
$u=0.45$ and $u=0.7$ are associated with branches of roots along
which growth rates peak rapidly and then the branch terminates {[}such
behavior may be expected from the multiple resonant denominators of
Eq. (\ref{eq:DR-cold}){]}.

\subsection{Warm-beam analysis}

\begin{figure}
\begin{centering}
\includegraphics[width=0.8\columnwidth]{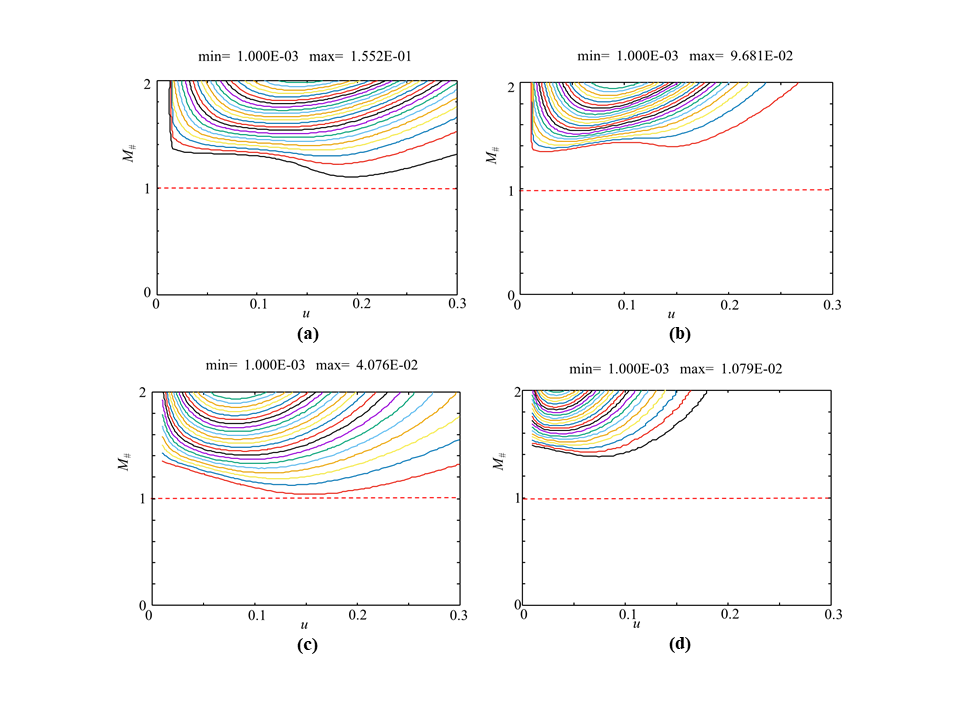}
\par\end{centering}
\caption{\label{fig:Wcons}Contours of warm-beam growth rate \emph{vs.} $u$
and Mach number $M_{\#}$ for four cases: (a) $m=1$ (linear spline)
without filter; (b) $m=2$ (quadratic spline) without filter; (c)
$m=1$ with filter; (d) $m=2$ with filter. Dashed horizontal line
indicates Mach 1.0 to gauge stability limit.}
\end{figure}

The effect of introducing a small beam thermal spread is examined
next by finding roots of Eq. \ref{eq:DR2} as previously described.
Figure \ref{fig:Wcons} shows the results of adding a beam temperature
to the previous results. The figures depict the growth rate as a function
of Mach number (defined in normalized units as $M_{\#}=u/\lambda$)
and drift velocity $u$. As can be seen, unstable modes exist only
for a drift velocity exceeding a given Mach number, which is essentially
1.0 for the various cases. While the maximum growth rate is quite
strongly dependent on the particulars of the method, the stability
limits are largely independent of these.

\begin{figure}
\begin{centering}
\includegraphics[width=0.7\columnwidth]{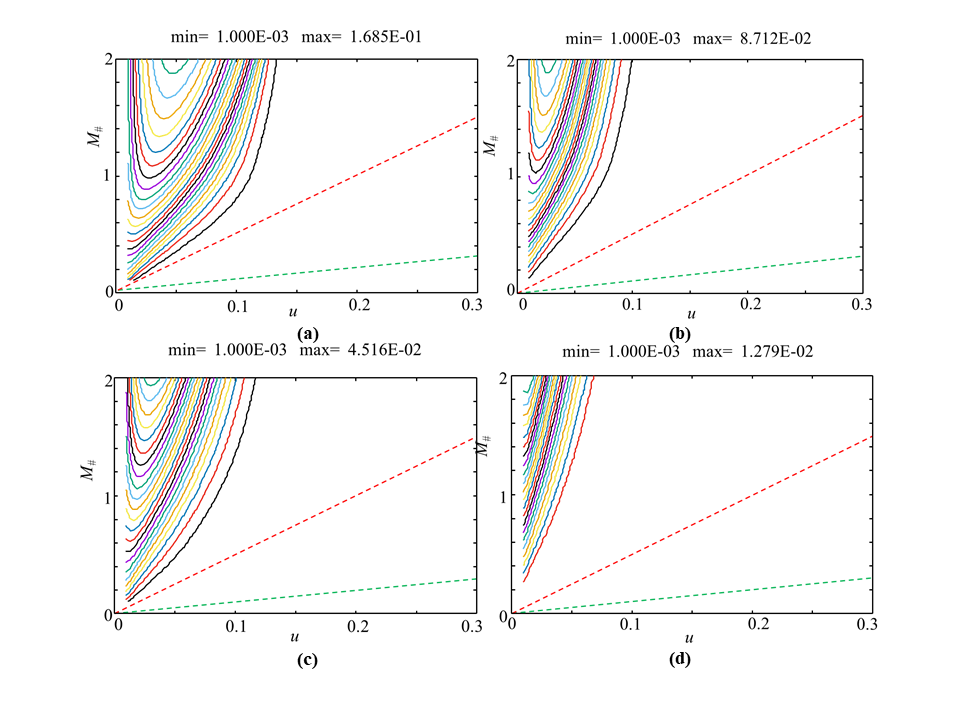}
\par\end{centering}
\caption{\label{fig:Pcons}Same plots as those of Fig. \ref{fig:Wcons} for
the momentum-conserving method. Dashed lines indicate $\Delta/\lambda_{D}=1/\lambda$
of 1 (green) and 5 (red).}
\end{figure}

These results for the energy-conserving formulation may be compared
with those for the usual momentum-conserving scheme \citep{birdsall-langdon}
employed in many PIC implementations. Figure \ref{fig:Pcons} shows
the same scan presented in Fig. \ref{fig:Wcons}, but repeated for
the momentum-conserving methods. The stable region at low Mach is
now replaced by the familiar cell-size to Debye length ratio stability
constraint known for the aliasing stability of momentum-conserving
PIC.

With the improved viewpoint provided by these wide-ranging scans,
we find a new result that the stability of EC-PIC for a warm-drifting
beam distribution is qualitatively different from that of usual momentum-conserving
PIC algorithms, which typically require cell sizes limited to a small
multiple of Debye length (Fig. \ref{fig:Pcons}). Namely, for energy-conserving
PIC, a new stability region appears at low to modest Mach numbers
that is independent of cell size. This result implies that, at worst,
finite-grid instabilities in energy-conserving PIC saturate when the
thermal spread of the beam becomes comparable to the drift velocity.
However, more importantly and as will be discussed subsequently, this
property allows the application of the algorithm to a wide range of
interesting ambipolar plasmas without interference from aliasing instabilities
in practice.

\begin{figure}
\begin{centering}
\includegraphics[scale=0.3]{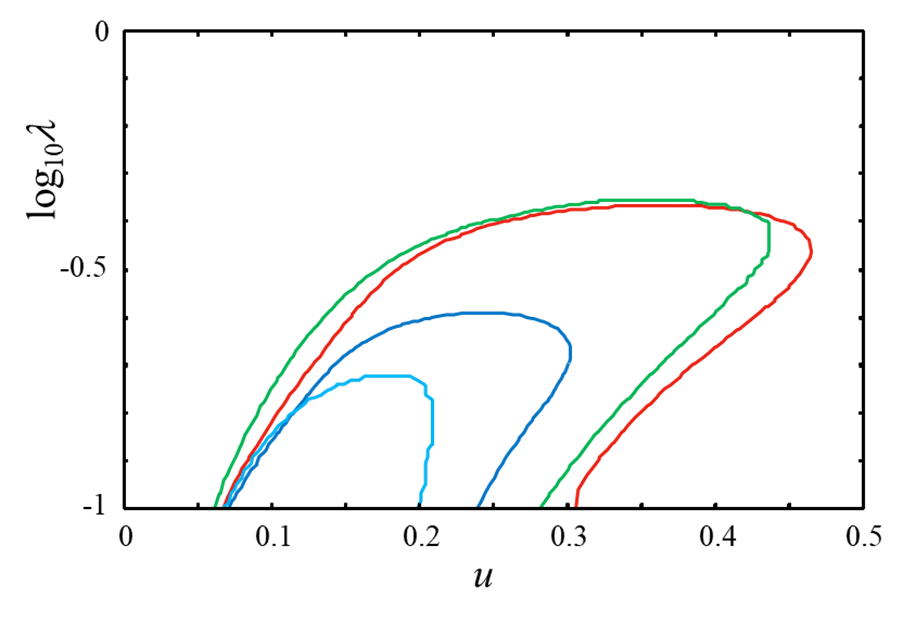}
\par\end{centering}
\caption{\textcolor{red}{\label{fig:absolute-stability}}Marginal contours
in $\log_{10}\lambda,u$ space for four cases: red -- linear spline,
no filter; green -- linear spline, filter; blue -- quadratic spline,
no filter; cyan -- quadratic spline, filter}
\end{figure}
It is also instructive to plot the stability diagrams for EC-PIC in
terms of the normalized Debye length $\lambda$ and drift velocity
$u$ (Fig. \ref{fig:absolute-stability}), to contrast against similar
diagrams for the momentum-conserving PIC algorithms. From this figure,
it is apparent that absolute stability is attained for arbitrary $u$
above a threshold in $\lambda=\lambda_{D}/\Delta$, i.e., when the
Debye length is sufficiently resolved. The threshold, however, is
strongly dependent on the interpolation order $m$ and the use of
filtering. While the impact of filtering for first-order splines is
minimal, it is significant for higher-order interpolations. For first-order
interpolation, the stability limit is $\Delta\lesssim4\lambda_{D}$.
For second-order splines, absolute stability with smoothing is achieved
for $\Delta\lesssim8\lambda_{D}$, which is a factor of 2 larger than
the first-order-spline stability limit (which would result in efficiency
gains of $2^{4}\sim16$ in 3D+time just from the easement of spatial
and temporal resolution requirements). For filtered third-order splines
(not shown), the limit is $\Delta\lesssim12\lambda_{D}$, resulting
in potential efficiency gains vs. first-order interpolation of more
than $1000$. The impact of smoothing on the stability thresholds
in $u$ is also apparent from Fig. \ref{fig:absolute-stability},
with the unstable region shrinking significantly with both spline
order and filtering.

\section{Numerical experiments}

\label{sec:numerical_exp}

We perform numerical simulations with the CCB algorithm ($m=2$) with
and without binomial smoothing, to confirm the validity of the previous
linear theory analysis. Our simulation setup considers $\Delta=1$
and $\omega_{p}=1$, and therefore code units correspond to dimensionless
units in the previous section. We consider a single-species simulation
with 256 particles per cell, and 128 cells (and therefore the domain
length is $L=128$). Both cold-beam and warm-beam simulations are
initialized with random particle positioning in the simulation coordinate
(to perturb all wavenumbers). Growth rates are obtained by Fourier-decomposing
the signal, and searching for the fastest growing mode. Since growing
aliasing modes are oscillatory, and saturation amplitudes for some
of these modes are small, for some of these measurements there is
a limited growth time window to obtain growth rates, and therefore
the growth rate measurements are only approximate.

\subsection{Cold-beam results}

\begin{figure}
\begin{centering}
\includegraphics[width=0.48\columnwidth]{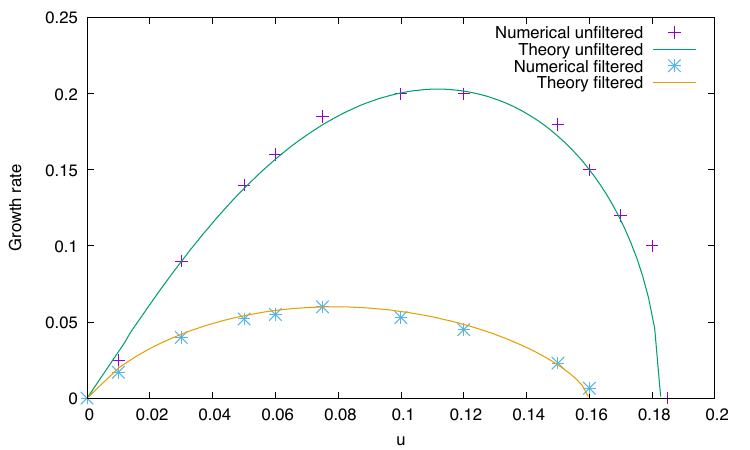}
\par\end{centering}
\caption{\label{fig:gamma-verif}Growth rate comparison between the cold-beam
theory and numerical experiment for $m=2$, with and without filtering. }
\end{figure}
We begin by verifying our cold-beam semi-analytical results with
numerical experiments. The results are depicted in Fig. \ref{fig:gamma-verif},
where numerical growth rates are compared against analytical growth
rates for the case of $m=2$, with and without binomial smoothing.
Agreement is excellent.

\begin{figure}
\begin{centering}
\subfloat[\label{fig:x-v-no-binom}Without filter at time 100. ]{\begin{centering}
\includegraphics[width=0.48\columnwidth]{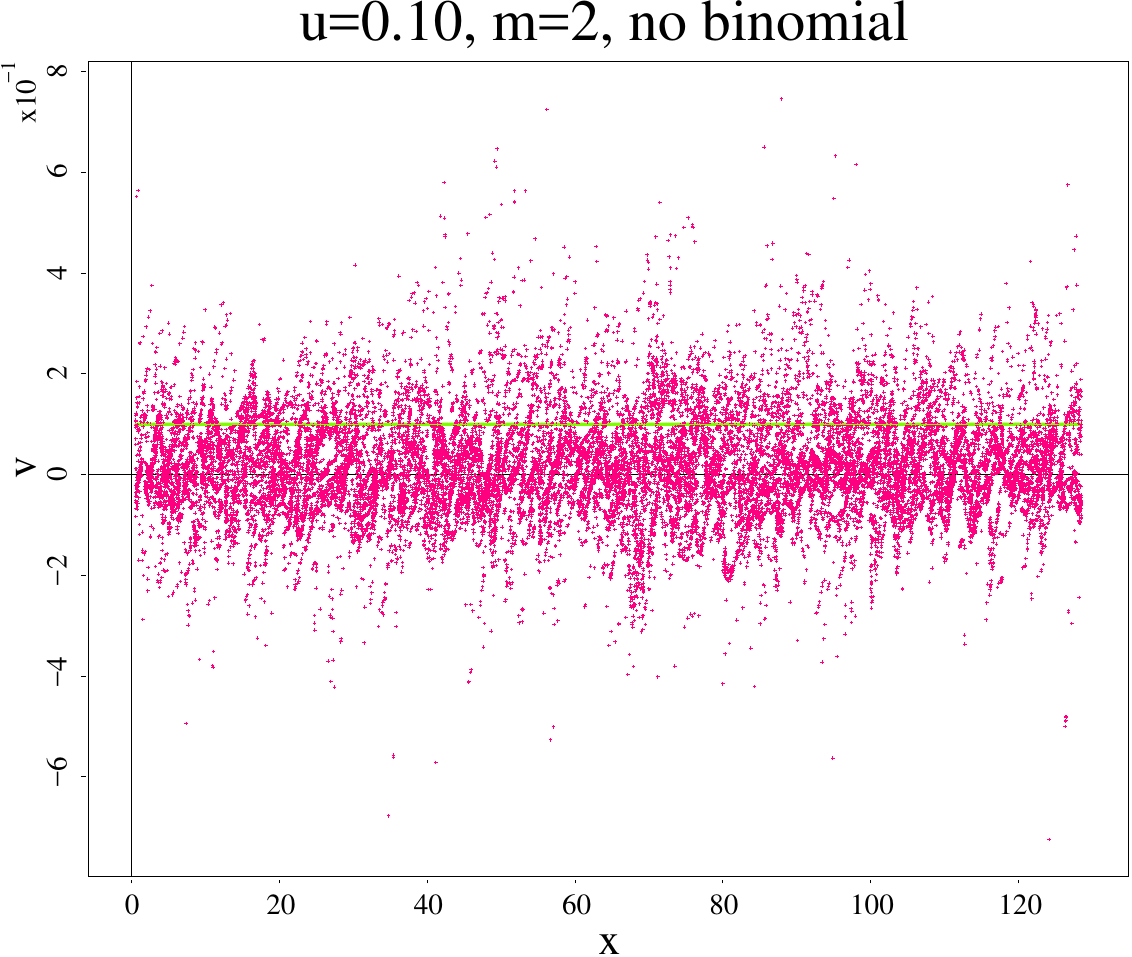}
\par\end{centering}
}\subfloat[\label{fig:x-v-binom}With filter at time 1000.]{\begin{centering}
\includegraphics[width=0.48\columnwidth]{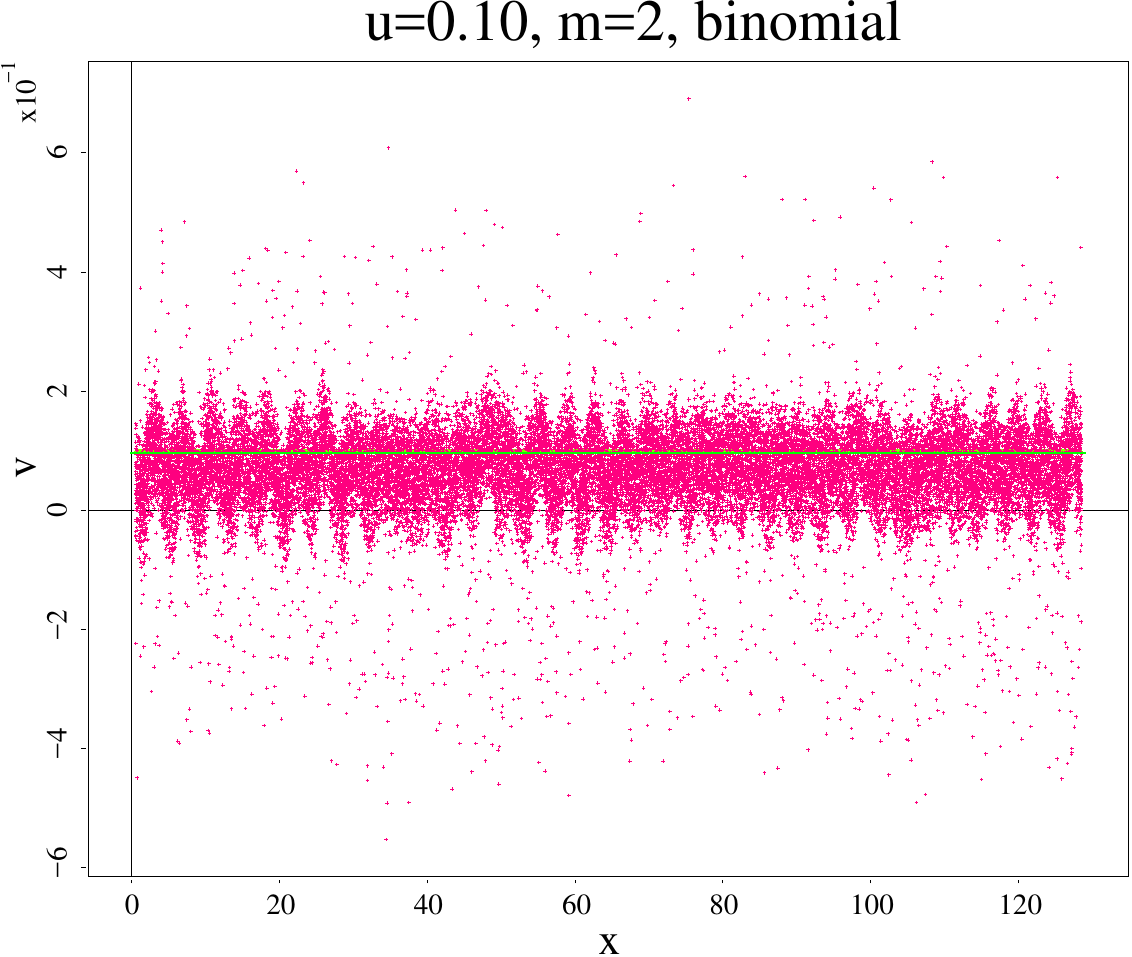}
\par\end{centering}
}
\par\end{centering}
\caption{\label{fig:x-v-plots}Nonlinearly saturated cold-beam phase-space
plots of initial (green) and final (red) states for $u=0.1$ and $m=2$,
with and without filtering.}
\end{figure}
Figure \ref{fig:x-v-plots} shows phase-space plots of initial and
final phase-space solutions of a cold-beam simulation with $u=0.1$,
$m=2$, with and without filtering ($t=100$ for unfiltered, 1000
for filtered). Charge and energy are conserved to round-off in both
cases, and the instability has saturated nonlinearly by the end of
the simulation. It can be clearly appreciated in both cases that the
instability saturates when the thermal velocity of the beam becomes
comparable to the initial drift velocity, which is consistent with
the warm-beam linear theory above. However, the filtered simulation
takes much longer to saturate, owing to the much smaller linear growth
rates. 

\subsection{Warm-beam results}

\begin{figure}
\begin{centering}
\subfloat[Unfiltered case ($u=0.1)$]{\includegraphics[width=0.5\columnwidth]{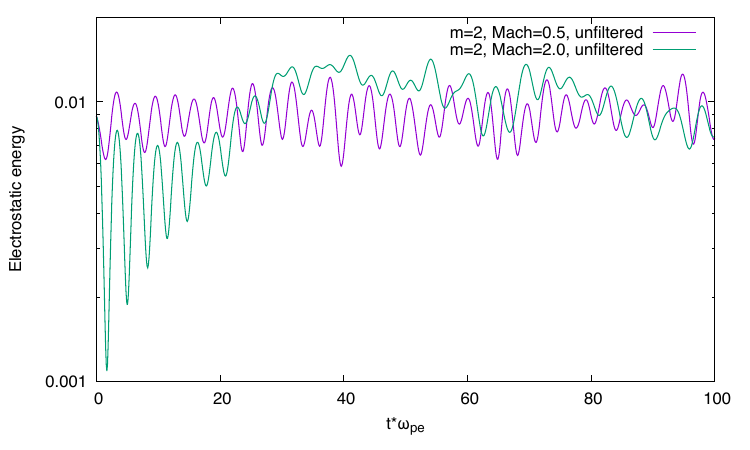}

}\subfloat[Filtered case ($u=0.05)$]{\includegraphics[width=0.5\columnwidth]{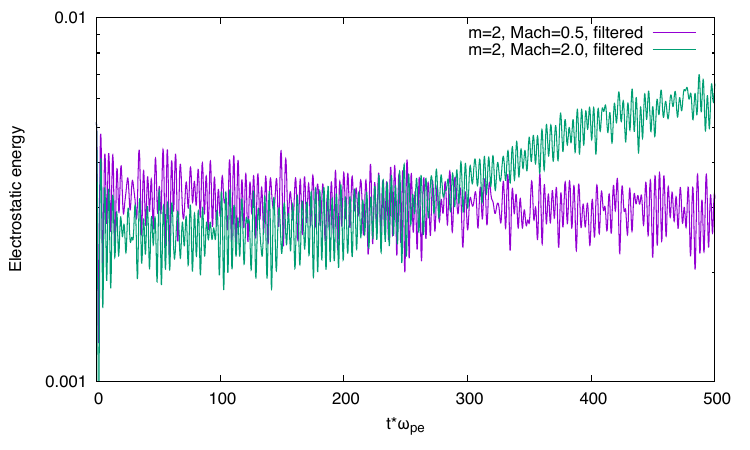}

}
\par\end{centering}
\caption{\label{fig:Warm-beam-numerical}Warm beam numerical time histories
for the electrostatic energy for $m=2$, both filtered and unfiltered.
The unfiltered case is more unstable, as evidenced by the much quicker
saturation of the instability.}
\end{figure}
Next, we test the linear-theory predictions of Fig. \ref{fig:Wcons}
for the warm beam with CCB. We have run both filtered and unfiltered
cases for $m=2$, with normalized drift velocities chosen to maximize
the growth rate ($u=0.1$ for the unfiltered case, $u=0.05$ for the
filtered one). We have chosen two Mach numbers, $M_{\#}=0.5$ (which
should be stable for both cases), and $M_{\#}=2$ (which should be
unstable and with sufficiently large growth rates, but much more so
for the unfiltered case). The results are presented in Fig. \ref{fig:Warm-beam-numerical},
which depicts time histories of the electrostatic energy, and confirm
the linear-theory conclusions. A detailed Fourier analysis of the
signal quantitatively confirms the growth rates computed by the linear
theory analysis and reported in Fig. \ref{fig:Wcons}.

\subsection{Plasma slab expansion into vacuum}

We conclude our numerical exploration of the stability properties
of EC-PIC with a 1D-1V plasma-expansion problem of a slab into vacuum.
This problem has been chosen to exemplify the constraints that ambipolarity
imposes on the electron Mach number, which leads to robust behavior
against aliasing instabilities even when $\lambda\ll1$. Our problem
setup is as follows. Normalization of time and space is chosen as
in the theoretical development presented earlier in this study. The
plasma slab has initial charge density of unity, and is localized
in a region of 25 Debye lengths at the right boundary. Ions and electrons
are in thermal equilibrium and at rest, with the electron thermal
velocity being unity (consistently with our normalization). We employ
realistic ion-to-electron mass ratio, $m_{i}/m_{e}=1846$. We simulate
a domain of 500 Debye lengths, employing a 64-point variable mesh
packed at the right boundary (with resolution of 1 Debye length),
and extending out to the left boundary (with resolution of 20 Debye
lengths). This mesh extension implies that the mesh is getting coarser
as the plasma cools and expands, thereby stressing the algorithm more.
The right boundary is reflecting, and the left boundary is open to
vacuum. We use 5000 particles per cell for both ions and electrons.
We use a second-order interpolation strategy ($m=2$) without filtering.
We run the problem to a final time of $T=5000\omega_{pe}^{-1}$.

\begin{figure}
\begin{centering}
\includegraphics[width=0.8\columnwidth]{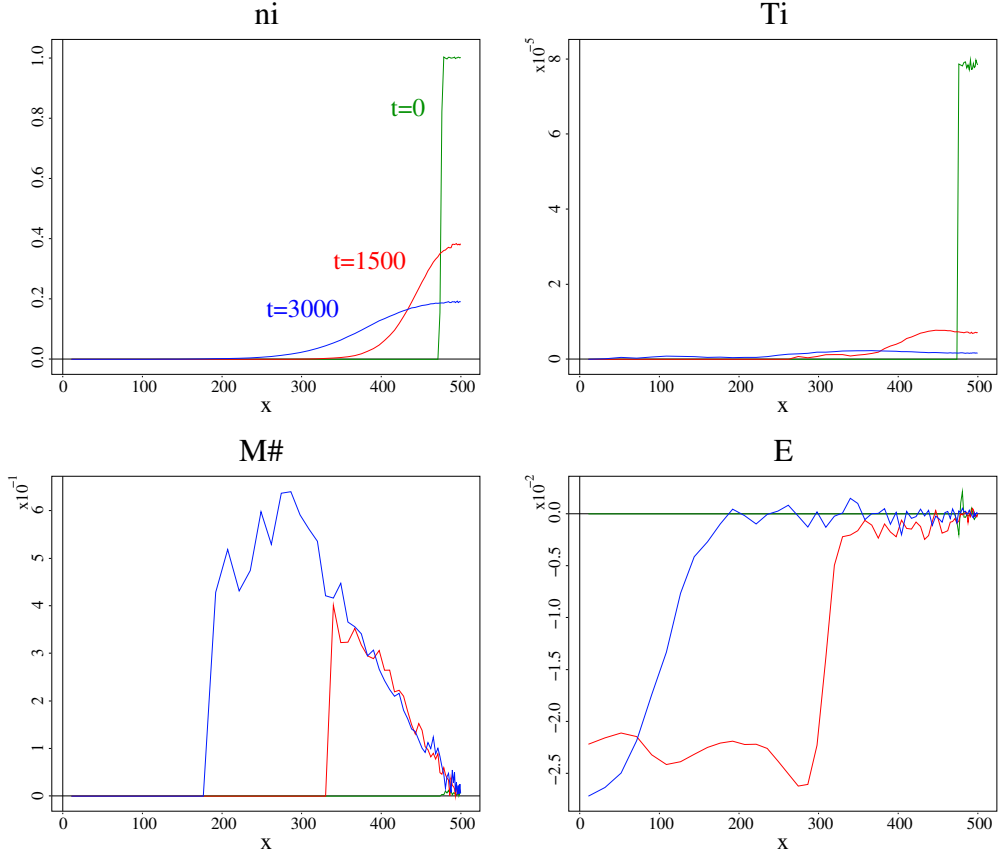}
\par\end{centering}
\caption{\label{fig:Plasma-exp-spatial-profiles}Spatial profiles of ion density
$n_{i}$, ion temperature $T_{i}$, electron Mach number $M_{\#}$,
and electric field $E$ at time $t=0$ (green), $t=1500\omega_{pe}^{-1}$
(red), and $t=3000\omega_{pe}^{-1}$ (blue), for the slab plasma expansion
problem. }
\end{figure}

Profiles of ion density, temperature, electron Mach number, and electric
field are depicted in Fig. \ref{fig:Plasma-exp-spatial-profiles},
where it can be clearly appreciated the development of a finite ambipolar
electric field as the slab expands due to faster electron motion.
Owing to the presence of the ambipolar electric field, electrons are
pulled back with the ions, and their Mach number is limited to about
$\sim0.3$, below the stability limit of the algorithm. The Mach number
diagnostic has been computed only when the ion plasma density is larger
than 1\% of the density at the right boundary at any given time, and
when the electron thermal velocity is larger than $10^{-4}$. This
prevents pollution of the diagnostic with regions that are either
very cold, or very rarefied, or both.

\begin{figure}
\begin{centering}
\includegraphics[width=0.7\columnwidth]{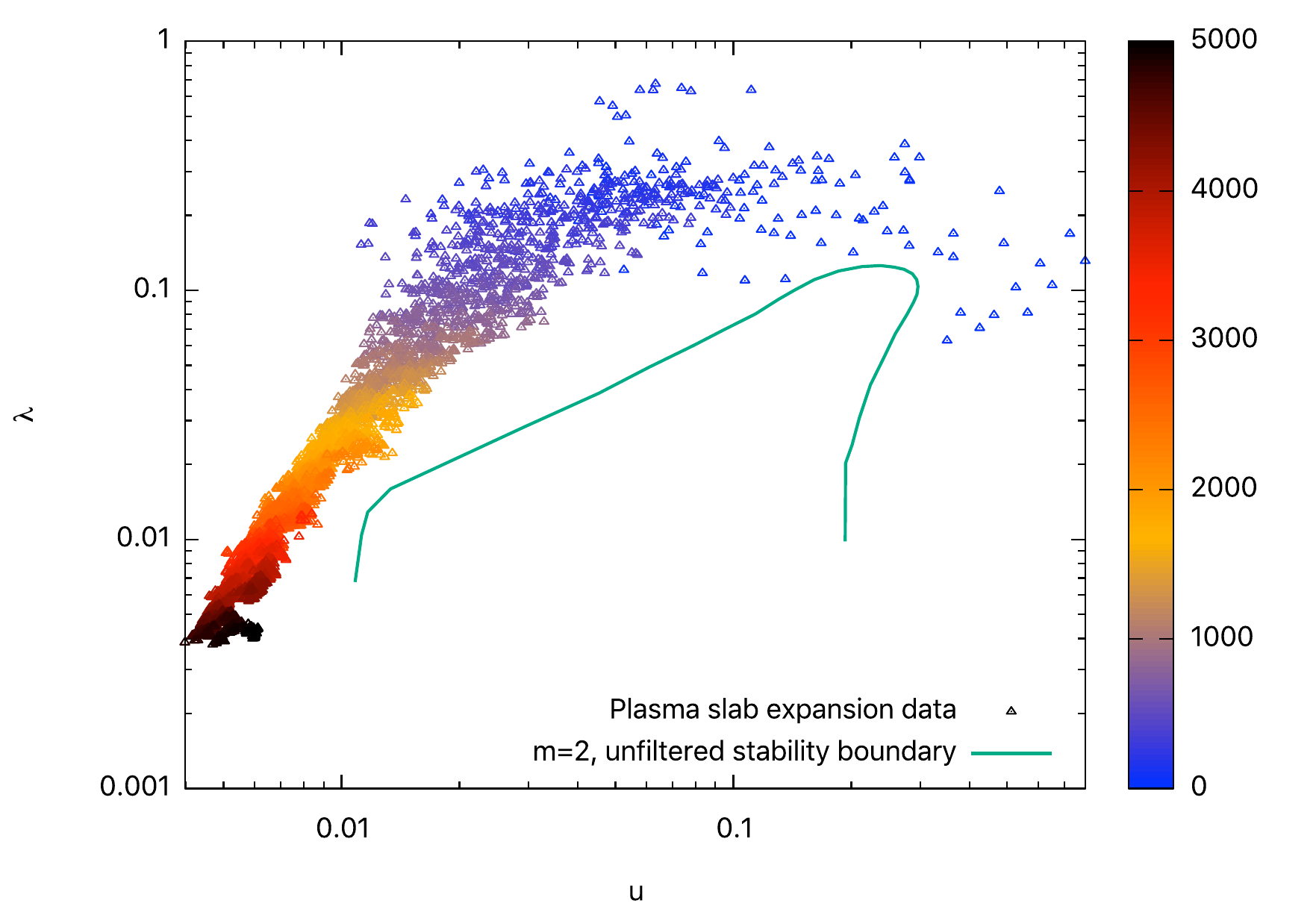}
\par\end{centering}
\begin{centering}
\includegraphics[width=0.7\columnwidth]{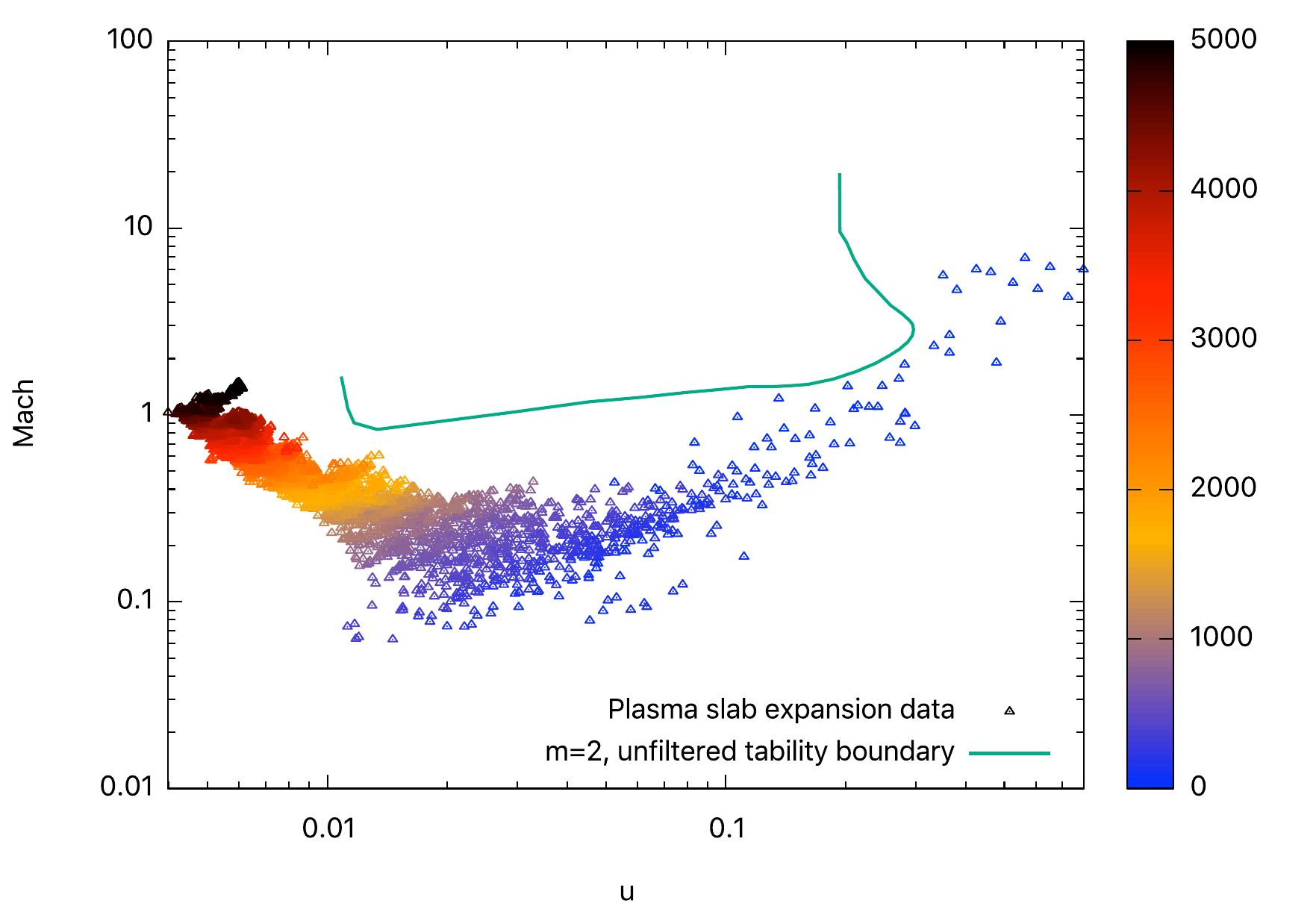}
\par\end{centering}
\caption{\label{fig:plasma-expansion-stability}Demonstration of ambipolar
stability in time of the unfiltered $m=2$ algorithm using the plasma
expansion problem. The green line represents the stability boundary
given in Fig. \ref{fig:absolute-stability}.}
\end{figure}
The strong stabilization of aliasing instabilities by the ambipolar
electric field is demonstrated in Fig. \ref{fig:plasma-expansion-stability},
where time-history plots of $u$-$\lambda$ and $u$-$M_{\#}$ pairs
for the plasma expansion problem are shown together with the stability
boundary for the unfiltered $m=2$ case presented in Fig. \ref{fig:Plasma-exp-spatial-profiles}.
The $u$-$\lambda$ (or $M_{\#}$) pairs are obtained as follows:
for each time step, we compute the place where the maximum Mach number
is located, and there we measure the normalized drift $u$ and normalized
Debye length $\lambda$. Different coloring of these pairs indicate
time, according to the color legend on the right of each plot. The
figure demonstrates that the expansion is concentrated in the stable
region of the parameter space, giving a perfectly stable evolution
despite $\lambda$ values significantly smaller than 0.1.

\section{Discussion and Conclusions}

\label{sec:conclusions}

We have formulated the cold and warm beam electrostatic PIC dispersion
relations including finite-grid effects in the limit of vanishing
time step. Using improved numerical methods and considerable compute
time, we surveyed a wide range of dimensionless parameters for both
energy- and momentum-conserving PIC methods. Results indicate a qualitatively
different stability picture to that offered by previous studies. Momentum-conserving
methods require a cell size a few times Debye length to avoid aliasing
instabilities and associated rapid plasma heating. Energy-conserving
methods, on the other hand, provide a new stability window at low
to moderate Mach numbers. The stability threshold is around Mach 1.0,
depending little on details of the particle shape and/or mesh filtering.
These results have been confirmed qualitatively and quantitatively
by comparison with PIC simulations using the modern methods of Ref.
\citep{chen-jcp-11-ipic}, and by the plasma slab expansion problem.

The implications of this new stability region are considerable for
ambipolar plasmas (an important subset of applications, particularly
of implicit PIC methods). In these plasmas, electrons remain coupled
to the much more massive ions, which prevent them from reaching large
drift speeds. As a result, electrons drift with ion-acoustic or smaller
speeds, and the electron Mach number should remain well below the
aliasing instability threshold found in this study. The Mach number
threshold is found to be largely independent of the details of particle
shape and mesh filtering strategies, indicating that it should be
possible to capture this desirable operation in a variety of practical
PIC methods (a confirmation of this result has been provided with
the plasma slab expansion problem). However, it depends strongly on
the ion-to-electron mass ratio, and therefore the effect may disappear
if one uses unrealistically heavy electrons. We conclude that, once
energy-conservation is built into the scheme, aliasing instabilities
are effectively tamed for ambipolar plasmas, regardless of other algorithmic
details.

For applications in which charge separation physics is critical (e.g.,
when driven by external means such as in laser-plasma interactions),
aliasing instabilities may remain an issue in EC-PIC regardless of
energy conservation, and judicious algorithmic choices in regards
to smoothing and interpolation order may be critical to avoid their
deleterious effects. In particular, we have shown that the combined
use of higher-order interpolation and filtering may alleviate resolution
requirements enough to speed up simulations by two to three orders
of magnitude.

Finally, it is important to note that our analysis has been focused
on the arbitrarily small timestep limit, and thus applies to both
explicit and implicit (when using small timesteps) EC-PIC algorithms.
However, as noted and demonstrated numerically in Ref. \citep{brackbill-forslund},
large timesteps may significantly slow down aliasing instabilities
in implicit PIC algorithms (after all, large timesteps can be looked
at as a low-pass filter on allowable temporal modes). We leave the
analysis of the large-timestep case for future work.

\section*{Acknowledgments}

The authors would like to acknowledge useful conversations with G.
Chen. LC was sponsored by the Office of Applied Scientific Computing
Research (ASCR) of the US Department of Energy. This work was partially
performed under the auspices of the National Nuclear Security Administration
of the U.S. Department of Energy at Los Alamos National Laboratory,
managed by LANS, LLC under contract DE-AC52-06NA25396.\pagebreak{}

\bibliographystyle{ieeetr}
\bibliography{../kinetic,../numerics}

\end{document}